\documentclass[reprint,amsmath,amssymb,aps]{revtex4-1}

\usepackage{graphicx}
\usepackage{dcolumn}
\usepackage{bm}
\usepackage{amsmath}
\usepackage{amssymb}
\usepackage{amssymb,bm,mathrsfs,bbm,amscd}

\begin{document}

\preprint{APS/123-QED}

\title{Epidemic Variability in Hierarchical Geographical Networks with Human Activity Patterns}

\author{Zhi-Dan Zhao$^{1}$}
\author{Ying Liu$^{1}$}
\author{Ming Tang$^{1}$}%
\email{tangminghuang521@hotmail.com}
\affiliation{$^1$ Web Sciences Center, University of Electronic Science and Technology
of China, Chengdu 610054, People's Republic of China\\
}
\date{\today}

\begin{abstract}
Recently, some studies have revealed that non-Poissonian statistics of human behaviors stem from the hierarchical geographical network structure. On this view, we focus on epidemic spreading in the hierarchical geographical networks, and study how two distinct contact patterns (i. e., homogeneous time delay (HOTD) and heterogeneous time delay (HETD) associated with geographical distance) influence the spreading speed and the variability of outbreaks. We find that, compared with HOTD and null model, correlations between time delay and network hierarchy in HETD remarkably slow down epidemic spreading, and result in a upward cascading multi-modal phenomenon. Proportionately, the variability of outbreaks in HETD has the lower value, but several comparable peaks for a long time, which makes the long-term prediction of epidemic spreading hard. When a seed (i. e., the initial infected node) is from the high layers of networks, epidemic spreading is remarkably promoted. Interestingly, distinct trends of variabilities in two contact patterns emerge: high-layer seeds in HOTD result in the lower variabilities, the case of HETD is opposite. More importantly, the variabilities of high-layer seeds in HETD are much greater than that in HOTD, which implies the unpredictability of epidemic spreading in hierarchical geographical networks.

\begin{description}
\item[PACS numbers]
89.75.Hc, 87.23.Ge, 05.40.-a
\end{description}
\end{abstract}

\maketitle

\textbf{Since the discovery of non-Poissonian statistics of human behaviors such as human interaction activities and mobility trajectories, more and more scientists have been paying attention to the role of these patterns in epidemic spreading. Most recent research results showed that both time and space activity characteristics respectively have significant impacts on spreading dynamics. However, it is still unclear to us how the spatiotemporal characteristics affect the prevalence. Indeed, the time characteristics of human activities is closely related to the space characteristics. Recently, some studies have revealed that non-Poissonian statistics of human behaviors stem from the hierarchical geographical network structure, in which we investigate how the scale-free characteristic of human contact activities influences epidemic spreading. We find that, compared with homogeneous contact pattern and null model, correlations between time delay and network hierarchy can remarkably slow down epidemic spreading, and result in a upward cascading multi-modal phenomenon. More importantly, high-layer seeds arouse large variabilities, while low-layer seeds result in several comparable peaks of variabilities, which makes the prediction of epidemic spreading hard. This work provides us further understanding and new perspective in the effect of spatiotemporal characteristics of human activities on epidemic spreading.}

\section{\label{sec:level1}Introduction}

In modern society, the intrinsic mechanism of epidemic spreading is a noticeable issue. To understand the spatiotemporal patterns of epidemic spreading, the accurate mathematical models of epidemic spreading are used as the basic conceptual tools. There are various disease models like SIS (susceptible-infected-susceptible) and SIR (susceptible-infected-refractory)~\cite{Anderson_1992_OxfordUP,Herbert_2000_SIAMReview}. With the booming development of complex network theory~\cite{boccaletti2006complex}, epidemic spreading in complex networks has been strongly catching scientists' eyes~\cite{Moore_2000_PRE,Newman_2000_PRL,Newman_1999_PLA,Newman_1999_PRE,Moukarzel_1999_PRE,Eguiluz_2002_PRL,May_2001_PRE,
Romualdo_2003_WileyV,Gallos_2003_PhysicaA,Boguna_2003_PRL,Moreno_2003_PRE,Liu_2005_EPL,Huang_2007_JSM,Grabowski_2004_PRE,Motter_2003_PRE,
Ravasz_2002_Science,Ravasz_2003_PRE,Noh_2003_PRE,Noh_2004_PRE,Watts_2002_Science,Watts_2005_PNAS,YooK_2002_PNAS,Warren_2002_PRE,
Gastner_2006_EPJB,Nemeth_2003_PRE,Huang_2006_PRE,Crepy_2006_PRE,Zhou_2006_PRE}. Most studies focus on the effect of network structures on spreading dynamics, including the small world property~\cite{Moore_2000_PRE,Newman_2000_PRL,Newman_1999_PLA,Newman_1999_PRE,Moukarzel_1999_PRE}, the scale-free property~\cite{Eguiluz_2002_PRL,May_2001_PRE,Romualdo_2003_WileyV,Gallos_2003_PhysicaA,Boguna_2003_PRL,Moreno_2003_PRE}, the community structure~\cite{Liu_2005_EPL,Huang_2007_JSM}, and the hierarchical structure~\cite{Grabowski_2004_PRE,Motter_2003_PRE,Ravasz_2002_Science,Ravasz_2003_PRE,Noh_2003_PRE,Noh_2004_PRE,Watts_2002_Science,Watts_2005_PNAS}, etc. Besides, both spatial distance~\cite{YooK_2002_PNAS,Warren_2002_PRE,Gastner_2006_EPJB,Nemeth_2003_PRE,Huang_2006_PRE,Crepy_2006_PRE} and contact capacity~\cite{Castellano_2006_prl,Zhou_2006_PRE} were found to have nontrivial impacts on epidemic spreading.

Since the discovery of non-Poissonian statistics of human behaviors such as human interaction activities~\cite{Barabasi_2005_nature} and human mobility trajectories~\cite{Brockmann_2006_Nature,Gonzalez_2008_Nature}, more and more scientists have focused on the role of these patterns in epidemic spreading. Most recent research results showed that both time and space activity characteristics respectively have significant impact on spreading dynamics. On one hand, the non-Poissonian nature of human interactions results in slow spreading in the long time limit~\cite{vazquez_2007_PRL,Iribarren_2009_PRL,yang_2011_physicaA}. An analytical prediction was then proposed to understand the emergence of the extremely long prevalence time in spreading dynamics~\cite{Min_2011_PRE}. Further investigation revealed that this phenomenon mainly stems from weight-topology correlations and the bursty activity patterns of individuals~\cite{Karsai_2011_pre}. By defining the dynamical strength of social ties, an interesting phenomenon was explained: although bursts hinder spreading at large scales, group conversations favor the local probability of propagation~\cite{Miritello_2011_pre}. On the other hand, the human mobility patterns have a significant influence on epidemic spreading~\cite{Ni_2009_PRE,TangM_2009_PRE,Tang_2009_EPL,WangP_2009_Science,Balcan_2011_NatureP,Belik_2011_prx}. As human traveling statistics follow the scaling law, the corresponding simulation results showed that the occurrence probability of global outbreaks is determined by human travel behavior~\cite{Ni_2009_PRE}. Considering two distinct individual mobility patterns (i. e., dynamical condensation and object traveling), both theoretical analysis and numerical simulations revealed that these patterns have an essential influence on epidemic spreading in scale-free networks~\cite{TangM_2009_PRE,Tang_2009_EPL}. In the study of the fundamental spreading patterns of mobile virus outbreak, Wang \emph{et al.} found that a bluetooth virus's spreading is constrained, which offers ample time for developing and deploying countermeasures~\cite{WangP_2009_Science}. Recent studies presented us that human mobility patterns are often dominated by specific locations and recurrent flows~\cite{Song2010b}. Balcan \emph{et al.}~\cite{Balcan_2011_NatureP} and Brockmann \emph{et al.}~\cite{Belik_2011_prx} thus studied contagion spreading in bi-directional movements, and showed that its dynamics is significantly different from random diffusive dynamics.

Although many researchers have studied the effects of time and space characteristics of human behaviors on prevalence, it is still unclear to us how the spatiotemporal characteristics affect the prevalence. Indeed, the time characteristics of human activities is closely related to the space characteristics. Take human contact activities for example. The spatial distance between individuals will inevitably lead to the time delay of their contact activities. In this point of view, more attention needs to be paid to the effect of spatiotemporal characteristics. Recently, some studies have revealed that non-Poissonian statistics of human behaviors stem from the hierarchical geographical network structure~\cite{Kalapala_2006_PRE,Han_2011_PRE}, in which we investigate how the scale-free characteristic of human contact activities influences epidemic spreading. We find that correlations between time delay and network hierarchy can significantly affect the spreading speed and the variability of outbreaks, and it is very difficult to accurately forecast epidemic spreading in the hierarchical geographical networks.

The paper is organized as follows. In Sec. II, we introduce the hierarchical geographical network model and the propagation processes in two contact patterns. In Sec. III, we investigate the effects of different contact patterns on epidemic spreading. Finally, we draw conclusions in Sec. IV.

\section{\label{Model Introduction}Model Introduction}
\subsection{\label{Hierarchical Geographical Network Model}Hierarchical Geographical Network Model}

\begin{figure}[b]
\includegraphics[width=6cm,height=6.5cm,angle=0]{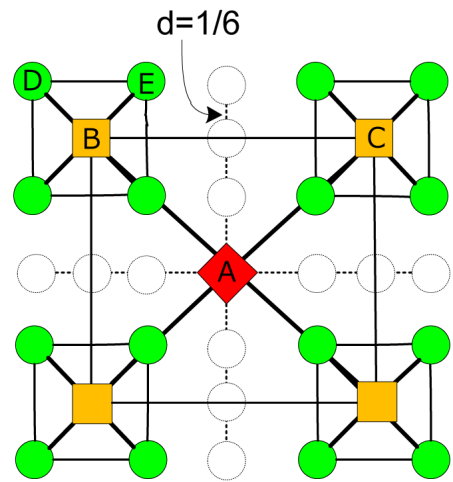}
\caption{\label{fig:hierarchyexapmple} (Color online) Illustration of the 2D hierarchical structure (L=3, M=4 and K=1) where "diamond", "squares", and "solid circles" represent the nodes in the \emph{1}st, \emph{2}nd and \emph{3}rd layer, respectively. The bolder line of each node represents links to its father node and the same layer node on its diagonal. Note that "hollow circles" denote no node is in these locations.}
\end{figure}

In order to reproduce the scaling law in human trajectories, a hierarchical geographical network model has been proposed~\cite{Han_2011_PRE}. In this model, all nodes are organized in $L$ layers. Denote $K$ as the number of first-layer nodes, and $M$ as the branching number of the current-layer nodes. Each of $l$th-layer nodes is connected to its father node, and two $l$th layer node are connected if they have the same father node. In the two dimensional plane, the whole area is divided into $K$ sub-regions, and $K$ 1st-layer nodes are assigned to locate in the center of them. Then, each of the $K$ sub-regions is further divided into $M$ sub-sub-regions, with the $KM$ 2nd-layer nodes locating in the center. Repeating this process until the $L$th-layer nodes are generated. Remarkably, there are strong correlations between the geographical distance and the network hierarchy in this network: the higher layer a node locates in, the farther it is from its father node and the other nodes in the same layer. In Fig.~\ref{fig:hierarchyexapmple}, a schematic diagram with $L=3, M=4, K=1$ is shown. $d_{BA}=\sqrt{2}/3$ (the distance between the second layer node $B$ and its father node $A$) is twice of $d_{DB}=\sqrt{2}/6$ (the distance between the third layer node $D$ and its father node $B$), and $d_{BC}=2/3$ in the second layer is twice of $d_{DE}=1/3$ in the third layer.

When a random walker continuously jump in this network, a power-law-like travel displacement distribution~$P(d)\sim d^{-2.5}$~is spontaneously generated in the thermodynamic limit~$t\rightarrow\infty$, where~$d$~is the geometric distance of a random walker jumping at each time step~\cite{Han_2011_PRE}.

\subsection{\label{Time delay in Contact Process}Time delay in Contact Process}
In various transportation networks, such as railway networks and airline networks, the time delay of human contact is determined by the geometric distance between two neighbors, where time delay is defined as the time interval when message is transmitted from one to the other. However, in communication networks, such as Internet and telephone network, time delay of contact has no relation with the geometric distance. Thus, contact patterns of human activities can be divided into two categories. The first pattern is the heterogeneous time delay (HETD) of contact activities, which depends on the geometric distance. For example, cities far from disease origins are infected later than those near the origins. Thus, the time delay of a contact~$\tau_{ij}$~is proportional to the geometric distance~$d_{ij}$~between two neighbors. Obviously, the time delay distribution of contact follows power law, which is consistent with the distribution of time intervals between two successive messages arriving at a given receiver~\cite{pica_2008_EPL}. The second pattern is the homogeneous time delay~(HOTD), in which the distance is ignored. For example, in communication networks, computer virus spreading is not restrained by the geometric distance. For this reason, we set the time delay~$\tau_{ij}$~of a contact a constant. To compare effects of different contact patterns on epidemic spreading, we set the mean time delay of all contacts as a fixed value in simulations. It is noted that time delay $\tau_{ij}$ is generally not an integer such as $1.3$. We separate its integral and decimal parts as $1.3=1+0.3$, and then reset the time delay $\tau_{ij}=1$ with probability $1-0.3$, while $\tau_{ij}=1+1$ with probability $0.3$. What can be imagined is that these two patterns can result in distinct spatiotemporal patterns of epidemic spreading.

\subsection{\label{Propagation Process}Propagation Process}

We study SI (susceptible-infected)~\cite{Anderson_1992_OxfordUP}~spreading dynamics in contact process~(CP)~through numerical simulations. In SI model, '\emph{S}' and '\emph{I}' represents respectively the susceptible (healthy) state and the infected state. At each time step of contact process, each infected node randomly contacts one of its neighbors, and then the contacted neighboring node will be infected with probability $\lambda$ if it is in the healthy state, or else it will retain its state. To eliminate the stochastic effect of the disease transmission, we set $\lambda=1$. In simulations, the propagation processes are as follows: (i) Select a node as the initial infected (i. e., seed) and all other nodes are in \emph{S} state. (ii) At each time step, the infected node $i$ in the active state randomly select one of its susceptible neighbors $j$, and then contact node $j$ after time delay $\tau_{ij}$. During this period, node $i$ is inactive. (iii) After $\tau_{ij}$, node $j$ is infected by node $i$. Meanwhile, node $i$ is reactivated. (iv) The propagation processes will continue until all nodes are in \emph{I} state.

Based on the above process, we investigate how these two patterns influence spreading speed and variability of prevalence in CP in hierarchical geographical networks. The prevalence is defined as the density of infected individuals $i(t)$ at time step $t$, and the spreading speed is defined as new case rate $n(t)$ at time step $t$, that is $n(t)=i(t)-i(t-1)$. With the spread of epidemic, new case rate increases to the maximal value~$n_{max}$, which denotes the occurence of outbreak. In order to analyze the impact of the underlying network topology
on the predictability of epidemic spreading, the variability of outbreaks is defined as the relative variation of the prevalence given by~\cite{Crepy_2006_PRE}

\begin{equation}\label{eq:variability}
\bigtriangleup[i(t)]=\frac{\sqrt{\langle i(t)^{2}\rangle-{\langle
i(t)\rangle}^{2}}}{\langle i(t)\rangle}.
\end{equation}
$\bigtriangleup[i(t)]=0$ denotes all independent dynamics
realizations are essentially the same, and the prevalence in the
network is deterministic. Larger $\bigtriangleup[i(t)]$ means
worse predictability that a particular realization is far from
average over all independent realizations.

\section{\label{sec:level2}The effect of different contact patterns}

\subsection{\label{Random seed}Random seed}
\begin{figure}[b]
\includegraphics[width=9.2cm,height=6.9cm,angle=0]{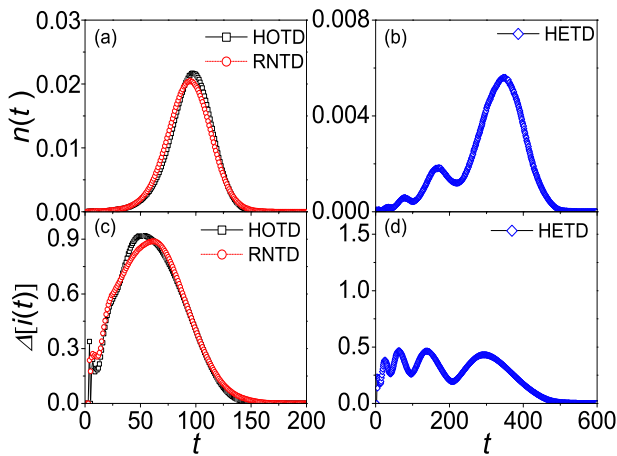}
\caption{\label{fig:newinfdes1} (Color online) The evolution of both $n(t)$ and $\Delta[i(t)]$ in the different contact patterns. $n(t)$ versus $t$ in HOTD ("squares")/RNTD ("circles") (a), and HETD ("diamonds") (b).  Correspondingly, $\Delta[i(t)]$ versus $t$ in HOTD/RNTD (c), and HETD (d). The parameters are chosen as $N = 9330, L=5, M = 6, K = 6$. The results are averaged over $10^{3}$ independent realizations.}
\end{figure}

In this paper, we study how the different contact patterns~(i. e., HOTD and HETD) influence epidemic spreading. We first pay attention to the case of random seed, that is a node selected randomly as the initial seed.  Obviously, in Figs.~\ref{fig:newinfdes1} (a) and (b), new cases in HETD has the lower peak value $n_{max}^{HETD}\approx0.006<n_{max}^{HOTD}\approx0.022$, and the longer full prevalence time $T_{f}^{HETD}\approx600>T_{f}^{HOTD}\approx175$, where the full prevalence time~$T_{f}$~is defined as the amount of time that all nodes of the network are infected by a seed. Meanwhile, a upward cascading multi-modal phenomenon in HETD is very intriguing, which is consistent with the results in Ref.~\cite{Watts_2005_PNAS,WangP_2009_Science,Grenfell_2001_Nature}. We suppose that this phenomenon might origin from the correlations between time delay and network hierarchy: time delays of contacts between the high-layer nodes are much greater due to longer distance from each other. To gain insight into the effect of these correlations, we employ null model where time delays are randomly exchanged between randomly chosen links~(RNTD), and the time delay-network hierarchy correlations are thus destroyed. In Fig.~\ref{fig:newinfdes1}~(a), $n(t)$ in RNTD displays almost the same unimodal pattern to that in HOTD, except for a slightly lower spreading (i. e., $n_{max}^{RNTD}\approx0.020<n_{max}^{HOTD}\approx0.022$). By comparison, we can conclude: although the heterogenous time delay of contact activities can result in slow spreading~\cite{vazquez_2007_PRL,Iribarren_2009_PRL,Min_2011_PRE}, the very low spreading is dominated by the time delay-network hierarchy correlations. More importantly, it indicates that these correlations certainly lead to the upward cascading multi-modal phenomenon. In this network, the outbreaks in low-layer sub-regions occur much earlier than that in high-layer sub-regions because low-layer nodes are much closer to each other. Even if the high-layer nodes are infected, outbreaks will not occur in a wider range until their child nodes are infected after a long time. Therefore, the peak values at $t\approx7,33,78,170,350$ in Fig.~\ref{fig:newinfdes1}~(b) correspond to the outbreaks in the \emph{5}th, \emph{4}th, \emph{3}rd, \emph{2}nd, and \emph{1}st layer sub-regions, respectively.

After that, we also investigate the variabilities of outbreaks in the different contact patterns. From Figs.~\ref{fig:newinfdes1}~$(c)$~and~$(d)$, we catch two essential differences between HETD and HOTD/RNTD: although the peak value of variability~$\Delta i_{max} \approx 0.45$ in HETD is one half of~$\Delta i_{max} \approx 0.90$ in HOTD/RNTD, there are four comparable peaks in HETD.  As pointed out in Ref.~\cite{Crepy_2006_PRE}, the variability changes coincide essentially with the evolution of the diversity of infected nodes. That is to say, the more homogeneously infected nodes distribute among all layers, the higher variability is. For the case of HOTD/RNTD, the variability is maximal when the diversity of layers of infected nodes is the largest at~$t\approx50/60$. In addition, there is a small fluctuation of the variability at~$t\approx7$ because of the rapid spread in the bottom-layer sub-regions~(i. e., a $M$ complete graph). For the case of HETD, the diversity of infected nodes does not vary monotonically over time, because there is always a maximum value of the diversity when outbreaks occur in the different layer sub-regions. As outbreaks successively occur from the low-layer to the high-layer sub-regions, the variability displays four comparable peaks. It means that the epidemic spreading in HETD has the variability over a long period of time, which brings a huge challenge to disease control.

\subsection{\label{Seeds from different layers}Seeds from different layers}
\begin{figure}[b]
\includegraphics[width=9.2cm,height=6.9cm,angle=0]{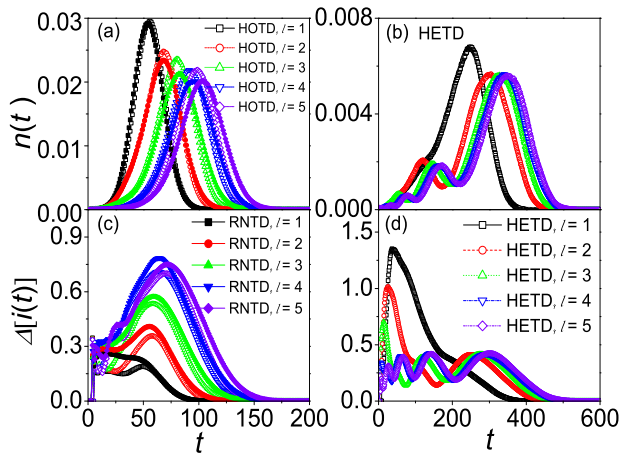}
\caption{\label{fig:newinfdes2}(Color online) The evolution of $n(t)$ and $\Delta[i(t)]$ for seeds from different layers where "squares", "circles", "triangleups", "triangledowns", and "diamonds" denote the cases of seeds from the \emph{1}st, \emph{2}nd, \emph{3}rd, \emph{4}th, and \emph{5}th layer, respectively. $n(t)$ versus $t$ in HOTD/RNTD (a), and HETD (b). $\Delta[i(t)]$ versus $t$ in HOTD/RNTD (c), and HETD (d). The results are averaged over $10^{3}$ independent realizations.}
\end{figure}

To confirm the dynamical centrality of nodes, it is very important to study the effect of different seeds on epidemic spreading~\cite{Kitsak_2010_nat,Gong_2011_chaos}. In this section, we investigate how seeds from different layers influence the spreading speed and the variability of outbreaks. In each simulation, a randomly chosen node in the designated layer is set as the seed. From Figs.~\ref{fig:newinfdes2} (a) and (b), $n(t)$ in three different contact patterns follow the same rule that the high-layer seeds can accelerate outbreaks. As mentioned in Sec. III. A, the only access to outbreaks in the wider range is across the high-layer nodes. Owing to time delay-network hierarchy correlations, the high-layer seeds doubtlessly make the outbreaks occur much fiercer, which corresponds to the larger peak value. It implies that central cities in transportation networks must be crucial regions for pandemic prevention and control.

Moreover, $n(t)$ in HETD displays an essential difference from the other two cases: the peak number of new cases rate is equal to the layer $l$ of the initial seed. For instance, there is single peak when a \emph{1}st layer node is set as the seed. On the one hand, owing to the characteristic of hierarchy, epidemic spreads from the high-layer nodes to the low-layer nodes, and thus the infected nodes gradually increase with time. On the other hand, the early prevalence is too slow due to time delay-network hierarchy correlations. After virus outbreaks in the the bottom layer, new cases rate rapidly decreases because of the effect of network size. Therefore, the seed from the $l=1$ layer results in only one peak. For the case of seeds from the $l=2$ layer, epidemic spreading from the seed to bottom layer leads to the first outbreak in the seed-centered sub-region, and the first peak value is greater than the new cases rate value for $l=1$~(i. e., $n_{l=2}\approx0.0021 > n_{l=1}\approx0.0018$)~at~$t=120$. Owing to the farther distance of the seed to both its father node and the same-layer nodes, the second peak corresponds to the outbreak in the wider range, but the second peak value $n_{max}^{l=2}\approx0.0056 < n_{max}^{l=1}\approx0.0068$. Similarly, new cases surely have $l$ peaks when the seed is in the \emph{l}th layer. These results have intriguing implications in hierarchical geographical networks: although the high-layer seeds make outbreaks occur much fiercer~(i. e., the higher peak), the low-layer seeds make outbreaks occur much earlier~(i. e., the earlier peak).

In Figs.~\ref{fig:newinfdes2}~$(c)$~and~$(d)$, it is a surprise that the outbreaks with the high-layer seeds in HOTD/RNTD have the lower variabilities, while the case in HETD is just the opposite. For the case of HOTD/RNTD, it takes almost the same amount of time to spread upward or downward due to no relationship between time delay and network hierarchy. When the seed is in the high-layer, epidemic can only spread downward, which has the less optional pathways. Thus, the high-layer seeds result in the less diversity due to the less optional pathways, which corresponds to the lower variability. As this contact pattern reflects the characteristics of information diffusion in communication networks, the above results demonstrate that information from the high-layer nodes have two obvious advantages: the faster diffusion rate and the better predictability.

For the case of HETD, it is very difficult for the virus to spread upward due to the longer distance from their farther nodes. On the contrary,
it's much easier to spread downward, which can arouse the more optional spreading pathways and the greater diversity. Thus, the outbreaks with the high-layer seeds have the higher variabilities. For example, $\bigtriangleup_{max}^{l=1}\approx1.35 > \bigtriangleup_{max}^{l=2}\approx1.00 > \bigtriangleup_{max}^{l=3}\approx0.70$. Intriguingly, the maximum value of the variabilities in HOTD/RNTD~$\bigtriangleup_{max}^{l=5}\approx0.71$~is only half of $\bigtriangleup_{max}^{l=1}\approx1.35$ in HETD. It indicates that the time delay-network hierarchy correlations make the predictability of epidemic spreading worse. On the other hand, although the outbreaks with the low-layer seeds have the better predictability, several comparable peaks of the variabilities, which corresponds to several outbreaks in the different ranges, cause a long-term trouble for pandemic prevention and control.

\begin{figure}[bh!]
\includegraphics[width=9.2cm,height=6.9cm,angle=0]{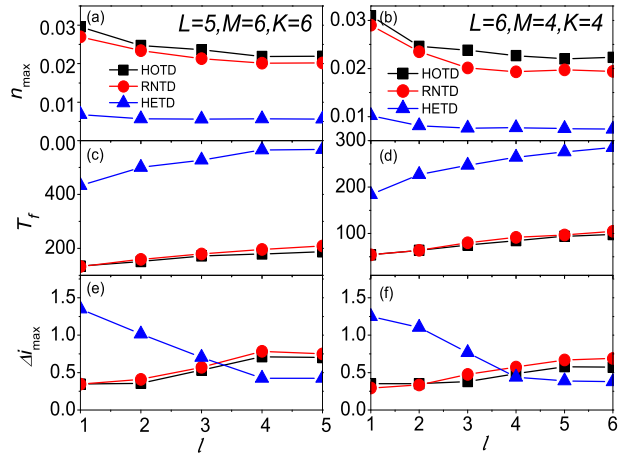}
\caption{\label{fig:genemopara} (Color online) In two networks with the different parameters, the prevalence and its variability as a function of the layer of seed~$l$~where "squares", "circles", and "triangles" denote the cases of HOTD, RNTD, and HETD, respectively. For $N=9330, L = 5, M = 6, K = 6$, the peak value of new cases rate $n_{max}$ (a), the full prevalence time $T_f$ (c), and the peak value of variability $\Delta i_{max}$ (e) versus $l$. For $N=5460, L = 6, M = 4, K = 4$, $n_{max}$ (b), $T_f$ (d), and $\Delta i_{max}$ (f) versus $l$.}
\end{figure}

In order to ensure the universality of the above results, other parameters are also chosen to simulate this process. Figs.~\ref{fig:genemopara} (b), (d), and (f) show the results for $L=6, M=4, K=4$. As expected, all simulations reveal the same conclusion: the trend of variabilities in HETD is completely contrary to that in HOTD/RNTD.
\section{\label{sec:level2}CONCLUSIONS AND DISCUSSIONS}
In conclusions, we have studied the effects of contact patterns on epidemic spreading in hierarchical geographical networks, and come to a clear understanding that different contact patterns (i. e., HOTD and HETD) can remarkably influence the spreading speed and the variability of outbreaks. First, we focus on the case of the random seed, and find that correlations between time delay and network hierarchy in HETD make epidemic spread much slower than HOTD/RNTD, and induce a upward cascading multi-modal phenomenon. Correspondingly, the variability in HETD is lower, but several comparable peaks make the long-term prediction of epidemic spreading hard. Second, we investigate the effect of seeds from different layers on epidemic spreading. For three contact patterns, the high-layer seeds make outbreaks occur much fiercer, while the low-layer seeds in HETD make outbreaks occur much earlier due to the small distance between nodes in the low-layer sub-regions. Interestingly, three contact patterns display distinct trends of variabilities with the different layers of seeds. The high-layer seeds in HOTD/RNTD result in the lower variabilities, the case of HETD is opposite. It is notable that the variabilities of the high-layer seeds in HETD are much greater than that in HOTD/RNTD, which implies the unpredictability of epidemic spreading in hierarchical geographical networks. To make matters worse, the variabilities of the low-layer seeds have more comparable peaks, which means it is difficult to accurately forecast epidemic spreading for a long time. This work provides us further understanding and new perspective in the effect of spatiotemporal characteristics of human activities on epidemic spreading.

\acknowledgments
Zhi-Dan Zhao would like to thank Tao Zhou, Duan-Bing Chen, Xiao-Pu Han, Yu Gan, and Jia-Bei He for stimulating discussions. This work is supported by the NNSF of China (Grants No. 11105025, 90924011, 60903073), China Postdoctoral Science Foundation (Grant No. 20110491705), the Specialized Research Fund for the Doctoral Program of Higher Education (Grant No. 20110185120021), and the Fundamental Research Funds for the Central Universities (Grant No. ZYGX2011J056).

\nocite{*}
\providecommand{\noopsort}[1]{}\providecommand{\singleletter}[1]{#1}%

\end{document}